\documentclass[twocolumn,nofootinbib]{revtex4}
\newcommand*\de{\mathrm{d}}
\renewcommand*\epsilon{\varepsilon}
\renewcommand*\phi{\varphi}
\renewcommand*\theta{\vartheta}
\def\beq{\begin{equation}}
\def\eeq{\end{equation}}
\def\bed{\begin{displaymath}}
\def\eed{\end{displaymath}}
\def\beqq{\begin{eqnarray}}
\def\eeqq{\end{eqnarray}}
\def\bedd{\begin{eqnarray*}}
\def\eedd{\end{eqnarray*}}
\usepackage{amssymb}
\begin{document} 
\title{VARIABLE MASS THEORIES OF GRAVITY}
\author{M. Leclerc}
\email{mleclerc@phys.uoa.gr}
\affiliation{Section of Astrophysics and Astronomy, Department of Physics\\
University of Athens\\ Panepistimoupolis, 157 84 Zografou}
\begin{abstract}
Several attempts to construct theories of gravity with variable mass 
are considered. The theoretical impacts of allowing the rest mass to vary 
with respect to time or to an appropriate curve parameter  
 are examined in the framework 
of Newtonian and Einsteinian gravity theories. In further steps, scalar-tensor
theories are examined with respect to their relation to the variation of the
 mass 
and in an ultimate step, an additional 
coordinate is introduced and its possible
relation to the mass is examined, yielding a five dimensional
space-time-matter theory.

\end{abstract}
\maketitle
\section{Introduction}
Many attempts have been proposed to generalize  Einstein's theory 
of gravity through the introduction of a variable rest mass. The motivation
for doing so comes mainly from theoretical arguments, like scale invariance 
of the gravitational theory, additional scalar fields 
that emerge from string theories or additional degrees of freedom  
that arise in the framework of brane-world theories \cite{ponce2}  and could 
eventually be related to such a mass variation. In these 
theories, the rest mass $m$ is supposed to vary slowly with time. 
Usually, the variation is considered to occur at a rate related to 
the age of the universe. In particular, this means that the effects of 
this departure from classical general relativity only occur at cosmological 
scales and do not affect planetary motion or any other experiment carried 
out in the planetary system. These effects have been studied in the framework
of several theories (see \cite{wesson} and references therein). 
In this article, we focus on the theoretical effects 
 of the assumption of  variable rest masses. Thus, instead of cosmological 
models, we consider the planetary system or even the gravitational field
of point particles and examine the impact of the introduction of the concept
of a variable rest mass. Even if there are no experimental changes, a
\textit{small} departure from general relativity may represent a 
\textit{severe}  
change in the theoretical concepts. Especially, attention should be paid 
to the self-consistency and the covariance of the theory as well as to 
the very definition of the mass as a quantity that characterises the 
particles.  

None of the considered models is essentially new and all of them has been 
considered with respect to several aspects in the literature. In this 
sense, the set of references is not complete and contains only those sources
which were explicitly consulted. The aim of this article is merely to 
drive attention to some of the problems that one should have in mind when 
dealing with variable mass theories.  
  
\section{Time dependent particle masses}
\subsection{Newtonian theory of gravity}
In a first attempt, we generalize Newton's theory of gravity by 
admitting a time dependence of the particle masses. The field equations
and the equations of motion read
\begin{eqnarray*}
\frac{\de}{\de t}\ (m(t)\vec v) &=& - m(t) \vec \nabla \phi\\
\Delta \phi &=& 4 \pi G \ \rho(\vec x,\ t),\\
\end{eqnarray*}
where $G$ is treated as a constant. 
The point particle case is now described by a time dependent 
mass density 
\begin{displaymath}
\rho(\vec x,\ t) = M(t) \delta(\vec x- \vec x(t)).
\end{displaymath}
A test particle of mass $m$ in the field of a central body at rest at the 
origin with mass $M$ evolves as 
\begin{displaymath}
\frac{\de }{\de t}\ (m(t)\vec v) = -\ \frac{Gm(t)M(t)}{r^3}\ \vec r.
\end{displaymath}
Although  the orbital angular momentum  $\vec L = m(\vec r \times \vec v)$ 
is still conserved, the observable kinematical quantity $\vec l = \vec r 
\times \vec v$ is not. We have instead
\begin{displaymath}
\frac{\de}{\de t}\vec l = - \frac{\dot m}{m} \vec l.
\end{displaymath}
The time dependence of the particle masses also destroys energy conservation.
For the case of a constant central particle mass, $M=M_0$, we find
\begin{displaymath}
\frac{\de}{\de t}\ (\frac{1}{2}\vec v^2 - \frac{GM_0}{r})= - \frac{\dot m}{m}
\vec v^2.
\end{displaymath} 
The Newtonian equations do neither tell us about the time evolution of
$m$ nore of $M$. In later sections, we try to remove this deficiency 
by treating the mass as a scalar field or as an additional dimension 
of spacetime, and thus as subject to the field equations. 

An interesting idea, which is the realization of Mach's principle in 
a very strong form, is to consider the rest mass of the particles as 
entirely due to its gravitational energy (or rather its opposite) from 
all the other surrounding
masses. Thus, if we suppose the energy to be of the Newtonian form, the 
mass could be expressed as (cf. \cite{booth})
\bed mc^2  = \sum{_M}{\frac{GmM}{r_M}},\eed
where the sum is carried out over all the bodies of the universe. Note that 
this is a recursive definition, since the other masses, $M$, also depend 
in the same way of $m$. Without going into details and discussing this 
equation, we will directly apply it to the case of planetary motion. 

Let $m$ be the mass of a planet, and $M$ the mass of the sun.   
Since the motion is confined at a small space region, we can suppose that 
the gravitational energy due to the interaction of $m$ or $M$ with all 
the other masses is  constant, and that the variation of the masses is
entirely due to the variation of the potential energy between $M$ and $m$. 

Thus, we can write
\bed mc^2 = m_0 c^2 + \frac{mMG}{r}= m_0c^2 +\frac{m_0M_0G}{c^2r}
+\frac{m_0^2M^2G^2}{c^2 r^2}+\ \dots 
\eed 
The (Newtonian) Lagrangian is given by
\bed L= \frac{m\vec v^2}{2} - m \phi, 
\eed 
with $\phi = -GM/r$. We now introduce our mass relation into this Lagrangian.
The second term can be expanded as
\bed \frac{mMG}{r}= (m_0+\frac{m_0M_0G}{c^2r})
(M_0+\frac{m_0M_0G}{c^2r})\frac{G}{r},
\eed
where terms of the order $\frac{1}{r^3}$ are neglected. 
This can be written as 
\bed \frac{mMG}{r} = -m_0 \phi +\frac{m_0^3}{c^2M_0}\phi^2 + 
\frac{m_0^2}{c^2} \phi^2, 
\eed
 where $\phi$ is now defined as $\phi = -\frac{M_0G}{r}$.
 As to the  first term, it  yields 
\bed \frac{m\vec v^2}{2}= \frac{m_0 \vec v^2}{2} 
+ \frac{\vec v^2}{2}\frac{m_0M_0G}{c^2r}= 
\frac{m_0\vec v^2}{2}-\frac{m_0\phi}{c^2 2}
\vec v^2,
\eed
where terms of the order $\phi\ \vec v^2/c^2$ are neglected 
(since the velocity 
$\vec v$ is nonrelativistic). 

The result is 
\bed L = \frac{m_0\vec v^2}{2} - m_0 \phi - \frac{m_0^2}{c^22}\phi \ \vec v^2 
+ \frac{m_0^2}{c^2} \phi^2 + \frac{m_0^3}{c^2M_0} \phi^2.
\eed
This is the Lagrangian that describes the motion of a planet $m_0$ 
for the considered mass variation, with an accuracy of 1 post-newtonian (1 PN) 
order. For planetary motion, the last term can of course be ignored, since
$M_0 >> m_0$. 

To the same order, the general relativistic result (from the post-newtonian 
expansion of 
the two body system) has the form (see \cite{landau}, \S 106) 
\bed L_{GR} = \frac{m_0\vec v^2}{2} - m_0 \phi - 
\frac{3m_0^2}{c^22}\phi\ \vec v^2 
- \frac{m_0^2}{c^2} \phi^2 - \frac{m_0^3}{c^2M_0} \phi^2, 
\eed
where, in the expression from \cite{landau},  
we have set the velocity of the central planet $M_0$ to zero.
(Actually, the going over from the 2 body problem to the one body problem, 
i.e. setting one  velocity to zero, should be accompanied by the 
relation $M>>m$, i.e. the last term should also be dropped. We will 
do so in the following.) 

As we have seen earlier in this section, the Newtonian energy is not 
a constant of motion any more. It is easy to see that the relation
\bed \frac{m_0\vec v^2}{2}+m_0 \phi = E_N \eed 
will suffer from first order corrections, i.e. we 
have $m_0 \vec v^2 /2+ m_0 \phi = E_N + O(1)$ 
This, however, leads to the following relation: 
\bed \frac{m_0^2\vec v^2}{c^22}\phi +\frac{m_0^2}{c^2} \phi^2  
= \frac{m_0}{c^2} E_N \phi + O(2)\eed 
Using this relation, we can bring our 1 PN order Lagrangian into the form 
\bedd  L &=& \frac{m_0\vec v^2}{2} - 
m_0 \phi - \frac{3m_0^2}{c^22}\phi\  \vec v^2 
- \frac{m_0^2}{c^2} \phi^2 + 2 \frac{m_0}{c^2} E_N\ \phi \\
&=& L_{GR} + 2 \frac{m_0}{c^2} E_N\ \phi.
\eedd 
In this form, the departure from general relativity is best seen. 
The last term is especially important for orbits near to the sun. 
This, of course, is a general feature of post-newtonian corrections. 
As the Lagrangian $L_{GR}$ describes already all the observable 
effects in the planetary system (except for the light deflection, 
which cannot be subject to a post-newtonian extension, which is 
confined to the description of particles with $\vec v << c$), it 
is surprising, that by our straightforward concept, we got the 
right form of the corrections (even if the factors don't match), using 
a non-relativistic scalar field theory. 

As a last remark to the Newtonian theory, we should have in mind that, even
if the above result would have been identical with the general 
relativistic expansion, the embedding of this theory into a special
relativistic context would again destroy this matching. Indeed, we would
have to bring the Newtonian equations into a covariant form, and the relation
of the mass to the gravitatial field would have to be changed. Problems arise, 
however, even before that!

\subsection{Special relativistic theory}
In a next step, we introduce the concept of a variable mass into 
a Lorentz covariant theory. To begin with, we consider the motion
of a test particle in flat spacetime under the influence of 
a electromagnetic field. The equations of motions in special relativity
follow from the  Lagrangian
\begin{displaymath}
L = \frac{1}{2}m\eta_{ik}u^iu^k +eA_iu^i.
\end{displaymath}
In order to conserve the Lorentz covariance, $m$ cannot be, a priori, 
a function of the time coordinate $t$ but should be regarded as dependent
on the curve parameter $\tau$ which is still to be interpreted. The
Euler-Lagrange equations yield
\bed 
\frac{\de}{\de \tau} \ (m(\tau)u^i)=eF^i_{\ k}u^k.
\eed
Contracting both sides with $mu_i$, we find a constant of motion $m^2u_iu^i$.
Thus, we can write down a relation 
\beq E^2 - c^2\vec{p}^{\;2} = m_0^{\;2}c^4. \eeq 
Energy and momentum are defined with $m(\tau)$ whereas the constant 
 $m_0$ is defined through the above relation. If we relate $\tau$ to 
 proper time, i.e. the time measured by a comoving observer, 
we find in the comoving frame ($u^{\alpha}=0,\ u^0=c$):
\bed m(\tau)=m_0. \eed
Thus, the nonmoving particle has a constant mass. This seems to be a 
reasonable relation, even though this interpretation of $\tau$ is not forced by
the equations of motion. In general, we only have $mu^0 = m_0c$ for the 
particle at rest. 

In terms of the observable kinematical quantity $u^i$, the equations of 
motion read
\bed \dot u^i = \frac{e}{m}F^i_{\ k}u^k - \frac{\dot m}{m}u^i.
\eed
The energy-momentum relation (1) allows us to eliminate $m$ and $\dot m$. 
The result is 
\bed \dot u^i = \frac{e}{m_0}\frac{u_ku^k}{c^2}F^i_{\ k}u^i + 
\frac{(u_lu^l)^{\dot{}}}{u_ku^k}u^i. \eed
However, it is now obvious that another constant of motion is $u_iu^i$, 
and hence $m(\tau)= m_0 = const$. The mass cannot be variable with the 
above Lagrangian! 

One may think of deriving the equations of motions from the following 
Lagrangian:
\bed L = \frac{1}{2}u_iu^i +\frac{e}{m(\tau)} A_iu^i.\eed
However, the corresponding equations of motion
\bed \dot u^k = \frac{e}{m}F^k_{\ i}u^i - \frac{e}{m}\frac{\dot m}{m}A^k\eed
break the gauge invariance of the electromagnetic theory for variable $m$,
which is not the scope of our theory. 

Finally, one may write down a parameter invariant Lagrangian of the form
\bed L_1 = m(\tau) \sqrt{\eta_{ik}u^iu^k} + e A_iu_i.\eed
This has the advantage of removing the ambiguity of whether we place 
the mass $m$ on the first term or its inverse on the second term.
Indeed, through the substitution $\de \lambda = \de \tau /  m$, the Lagrangian 
takes the form
\bed L_2 =    \sqrt{\eta_{ik}u^iu^k} +\frac{e}{m(\lambda)}A_iu^i,\eed
where $u^i$ is now defined as $\frac{\de x^i}{\de \lambda}$. 
Again, the second Lagrangian apparently breaks gauge invariance for 
variable $m$. As it is equivalent to the first one, we can immediatly 
conclude that $m$ is a constant of motion. Indeed, $L_1$ yields 
the equations
\bed \frac{\de}{\de \tau}(\frac{mu^i}{\sqrt{u_lu^l}})=e F^i_{\ k}u^k, \eed
which, contracted with $\frac{mu_i}{\sqrt{u_ku^k}}$ leads to $\frac{\de}
{\de \tau}m = 0$. 

The result of this section, namely that the mass of a test particle 
remains constant throughout its motion, holds true, of course, in the 
non-relativistic limit. The possibility of a variable mass obtained 
in the previous section is thus entirely due to the presence of a 
gravitational field, especially to the fact that the mass of the test particle
occurs at both sides of the equations of motion. Having this in mind, one 
should conclude that the variation of the mass should be subject not 
to the kinematical evolution of the particles, but to the underlying theory
of gravitation. Moreover, in the limit of vanishing gravitational fields, the
particle masses should be found to be constant.

\subsection{General relativity} 

The generalization to an arbitrary spacetime metric of the above 
considerations is straightforward. For a given metric, we find the 
equations of test particle motion by simply replacing the partial 
derivates through the covariant ones. The equation $\frac{\de m}{\de \tau}
=0$ however will not be affected! 

Next, consider the field equations of general relativity
\beq G_{ik} = \frac{8\pi G}{c^4}\ T_{ik}.\eeq
It is known that the only spherical symmetric vacuum solution is the
Schwarz\-schild solution, which can be written in the form
\bed \de s^2 = (1-\frac{R}{r})\ c^2\de t^2 - (1-\frac{R}{r})^{-1}\ \de r^2
- r^2\de \Omega^2, \eed 
where $R$ is related to the mass of the central planet through 
$R=\frac{2GM}{c^2}$. In solving the Einstein equations, $R$ arises 
as a constant of integration. The identification with the mass $M$ is 
usually made by considering the Newtonian limit. However, stricktly spoken,
the Schwarzschild solution is not really a \textit{vacuum} solution, but 
a solution corresponding to a stress-energy tensor of a point-like particle, 
i.e. 
\bed T^{ik}(\vec x,\ t) 
= \frac{M}{\sqrt{-g}}\ 
\frac{\de x^i}{\de \tau} \frac{\de x^k}{\de t}\ \delta^{(3)}
(\vec x - \vec x(\tau)), \eed
where $x^i(\tau)$ describes the worldline of the particle. 
In the comoving frame, the only equation  involving the matter fields  
reads, for the particle at the spatial origin,  
\bed G^0_0= \frac{8\pi G}{c^4}\ \frac{Mc^2}{\sqrt{-g}}\ 
\delta^{(3)}(\vec x). \eed
From this, we see that the constant of integration is already fixed by
the field equations. Now, we suppose that the central mass varies throughout
its evolution, i.e. $M=M(\tau)$. In the comoving frame,  the parameter
$\tau$ of the particle coincides with coordinate time $ct$. Thus we have
to solve Einstein's equations with a time dependent energy-stress tensor. 
 
In a first approximation, if we suppose that the time variation of $M$ 
is of negligible order (so that time derivates appearing in $G_{ik}$ can 
be neglected), we can write the solution as  
\bed \de s^2 = (1-\frac{2GM(t)}{c^2r})\ c^2\de t^2 - 
(1-\frac{2GM(t)}{c^2r})^{-1}\ \de r^2
- r^2\de \Omega^2. \eed 

It turns out, however, that the time dependent energy-stress tensor 
is not consistent with the underlying theory. Infact, 
as a consequence of the field equations (2), the matter also has to 
obey the relation 
\bed T^{ik}_{\ ;k} = 0. \eed 
For the point particle energy-stress tensor, this 
equation is equivalent to the equation of motion obtained from 
the Lagrangian 
\bed L = M(\tau)g_{ik}u^iu^i,
\eed 
and as we have seen, this leads to $M(\tau)= const$. (Note that the
we have already partly used the above equations when we choose the rest 
frame of the particle, i.e. when we set $u^{\alpha}= 0,\ \alpha=1,2,3$.)

As a conclusion to  this and the preceeding section, we can say that 
it is not possible, in a straightforward manner, to suppose that particle
masses vary throughout the evolution on their worldlines. The reason for 
this can be seen in the relation $E^2-c^2\vec p^{\;2}=const$ and its 
general relativistic generalization. In the Newtonian case, their is no
such relation, and the mass evolution is left, a priori, undetermined. 

\section{Scalar-tensor theories of gravity}
   
A quite different approach to a varying mass theory is obtained by 
considering the mass not as a function of the parameter of the 
worldline but as a spacetime dependent quantity, i.e. a scalar field
$m= m(x^i)$. In order to make contact to classical theories, it is 
convenient to write $m(x^i)=m_0e^{\phi(x^i)}$, where $m_0$ is a constant
which can be identified as the mass of a particle for vanishing gravitational
fields or the mass at some (cosmological) time $t_0$, depending on the 
model. 

The \textit{free} particle motion in flat spacetime is described by
\beq L = m_0 e^{\phi(x^i)}u_iu^i. \eeq
It is tempting to interprete $\phi(x^i)$ as the gravitational field
and to look for appropriate field equations. 
It is indeed possible to describe planetary motion up to Newtonian 
order through the identification $\frac{\phi}{2}= \phi_{_{newton}}$, 
but higher order corrections are not in agreement with experiment. Especially, 
 no deflection of light rays can be deduced from (3), in agreement 
with the fact that the vacuum Maxwell theory is conformally invariant. 

One should thus consider the scalar field as an \textit{additional} field. 
The simplest extension of general relativity containing a scalar field
is Brans-Dicke theory, which is based on the following action:
\begin{eqnarray*}
 S&=& (16\pi)^{-1} \int{\de^4 x \sqrt{-g}\ (\phi R- \omega \phi^{-1}
\phi_{,i}\phi^{,i})}\  \\&& + \int{\de^4 x \sqrt{-g} \mathcal L_m}. 
\end{eqnarray*}
 The resulting field equations read
\begin{eqnarray}
 G_{ik} &=& 8\pi \phi^{-1}T_{ik} + \omega \phi^{-2}(\phi_{,i}\phi_{,k}-
\frac{1}{2}g_{ik}\phi_{,l}\phi^{,l}) \nonumber
\\ && + \phi^{-1}(\phi_{;i;k}-g_{ik}\phi^{;l}_{;l}),
\end{eqnarray}
\beq  \phi_{;i}^{;i}= \frac{8\pi}{3+2\omega}\left[T 
- 2 \phi \frac{\partial T}
{\partial \phi}\right]. \eeq
These equations contain the following identity, which can be interpreted
as the equation of motion for the matter: 
\bed T^{ik}_{\ \ ;k} - \frac{\partial T}{\partial \phi}\ \phi^{,i} = 0. \eed

The presence of the scalar field
now allows one to write down an energy stress tensor for a particle 
with variable mass: 
\beq T^{ik} = \frac{m(\phi)}{\sqrt{-g}}\ \frac{\de x^i}{\de \tau}\frac{\de x^k}
{\de t} \delta^{(3)}(\vec x - \vec x(\tau)).
\eeq 
The field equations with this energy-stress tensor can be solved using a 
post-Newtonian extension \cite{estabrook}. 
Planetary system observations are in agreement 
with the theory for $\omega > 3000$, without taking into account the 
variation of the mass. The results of general relativity are found for 
$\omega \to \infty$ (identifying $G= \phi^{-1}(4+2\omega)/(3+2\omega)$).

The variation of the masses leads to further effects, none of which
have been observed up to now however. Generally, the scalar field, entering 
the equations of motions and the field equations in a different manner, 
may violate the weak equivalence principle. The equations of motion depend
on $m(\phi)$ and thus on the particle's structure. Of special interest is
the emission of dipole gravitational radiation from binary systems which is
not present in general relativity and could be subject to future experiments. 

It should however be emphasized that the variable mass entering the 
above energy-stress tensor is not a direct consequence of the theory. 
It should not be applied to elementary particles but only to extended 
bodies. It was actually introduced in the framework of hydrodynamics 
for Brans-Dicke theory (Nordtvedt effect, cf. \cite{will}).
The quantity $m$ varies because of the presence of the field 
$\phi$, which contributes to the internal energy of the body. The  
tensor (6) is just a convenient way to parametrize the matter by just one
field $m(\phi)$  and treat it as a point particle. A more physical 
interpretation is to use the identification $G= \phi^{-1}(4+2\omega)/(3+2
\omega)$ (see \cite{eardley} and references therein). 
The variation of $G$ then leads to the variation of the total 
energy which we described by $m$. 
 
Another remark is that the field equations do not determine the 
function $m(\phi)$. This is of course not surprising, since we 
have treated the body as a point particle. To find this function, 
one has to look for inner solutions of the body under investigation. 
It then turns out that the bodies can conveniently be parametrized 
by their \textit{sensitivitie} $s$, which is defined as
\bed s = - \ \frac{\partial(\ln m)}{\partial(\ln G)}, 
\eed 
with $G=\phi^{-1}(4+2\omega)/(3+2\omega)$ and the models 
give values between $s=0.1$ and $0,3$ for neutron stars
of masses around $1,4 M_{\circledcirc}$ 
and $s=0,5$ for black holes \cite{zaglauer}.   
  
The general remarks of this section hold true for every scalar-tensor 
theory based on general relativity. The scalar field can always be 
interpreted as a variation in the mass or as a variation in the 
gravitational constant, at least in a certain limit, i.e. if the 
departures from general relativity are supposed to be small (which 
has to be the case for the theory to be in agreement with experiment). 
The interpretation as variation of the gravitational constant is 
however more straightforward. Differences occur when we couple the 
theory to other fields and/or when we consider elementary particles, 
for instance the Dirac equation. In the latter case, a variable 
mass of the elementary particle would lead to many difficulties 
we would have to investigate and for which we do not have an
experimental justification (see section IV.F).  

\section{Variable mass through a fifth dimension}

\subsection{Introduction}

Finally, we take a look at theories who try to impose the concept 
of variable masses in general relativity through the use of a 
fifth dimension. The five dimensional space is called space-time-mass
and the metric is taken as (\cite{wesson}, \cite{gladush}) 
\begin{eqnarray} 
dS^2 = g_{AB}\de x^A\de x^B&=&
g_{ik}(x^m, x^4)\de x^i\de x^k \nonumber\\&&
+ \epsilon \phi^2(x^m, x^4)(\de x^4)^2, 
\end{eqnarray} 
where $\epsilon = \pm 1$, depending on which signature we choose\footnote{This
choice should rely on experimental data. Since no experiment is known that 
contradicts GR, the wisest choice would consist in setting $\epsilon=0$\dots}.
We take
the convention that capital indices $A, B,...$ run from 0 to 4 whereas
the indices $i,k,l...$ run from 0 to 3. The mixed components $g_{4i}$ 
have been transformed to zero through an appropriate coordinate
transformation. 

The action is taken to be 
\bed S = \int{R \sqrt{-g}\ \de^5 x} \eed 
and the resulting (vacuum) field equations read 
\beq G_{AB} = 0. \eeq
$R$ and $G_{AB}$ are constructed from the five dimensional metric 
 $g_{AB}$ in the usual way. 

\subsection{Spherically symmetric solution}

Before we interpret these equations and eventually generalize them 
to include matter fields, 
we begin our discussion by considering the following exact spherically 
  symmetric solution of (8)  
\beq \de S^2 = (1-\frac{a}{r})c^2\de t^2 - \frac{\de r^2}{1-a/r}-r^2
\de \Omega^2 +\epsilon (\de x^4)^2. 
\eeq
Here, $a$  arises as a  constant of integration. 

The idea of space-time-mass theory is to identify the fifth coordinate 
with the mass through the definition $x^4 = Gm/c^2$ (see \cite{wesson}). 
To be in 
agreement with experiment, the variation of the mass, $\de x^4$ has to 
be very small (at least for systems of dimensions of the solar system), 
and the above solution thus differs only slightly from the Schwarzschild 
metric if we identify $a$ with $2GM_0/c^2$, where $M_0$ is the mass of the 
central planet. One should however avoid to write $x_0^4 = GM_0/c^2$. A 
coordinate cannot be a constant of integration of the solution of the 
field equations. The solution has to be valid in all 5 dimensional 
spacetime (except 
for singularities), and thus for all $(x^m,\ x^4)$. In other words, the 
coordinate $x^4$ is related to the mass of a test particle in the above 
metric, and not to the (constant) mass of the source planet. The coordinate
$x^4$ is not constant, all that can be said is that the metric $g_{AB}$ 
does not depend on it. The quite different manners in which enter the 
masses of the test particle and of the source may eventually be removed 
if one considers the fully 2-body equations, which requires however 
that we solve the field equations with matter.

\subsection{Geodesic motion and matter fields}
Since we deal with a five dimensional theory, it is natural to require 
that test particles will follow five dimensional geodesics, which 
follow from the Lagrangian
\beq L = g_{AB} u^Au^B,
\eeq
where $u^A = \de x^A/\de S$. These equations can easily be reparametrized 
with the four dimensional parameter $\de s = c \de \tau$. The question
which parameter is the more physical one and what is their relation to 
proper time has been extensively discussed in the framework of Kaluza-Klein
theory (see \cite{ponce} for a detailed discussion). 
More fundamental however is the fact, that we cannot write 
instead of (10) 
\bed L_2  = m g_{AB} u^Au^B, 
\eed 
because $m$ being a coordinate, $L_2$ is not a scalar function. The question
thus arises how we shall describe a particle under the influence of both 
a gravitational and an electromagnetic field for instance. 
Since we know from general
relativity that the equations of motion follow directly from the field 
equations, we have to discuss the coupling of matter to the gravitational 
field in order to answer this question. 

\subsection{Induced matter theory}

There are two ways of introducing matter fields to the five dimensional 
theory. One is to complete the action with an adequate matter 
Lagrangian density, i.e. to write 
\bed S = \int (-\frac{c^4}{16\pi \Gamma}R + \mathcal L_m)\ \sqrt{-g}\ \de^5 x 
\eed 
 which leads to 
\beq G_{AB} = \frac{8\pi \Gamma}{c^4}\ T_{AB} \eeq
where $T_{AB}$ is the five dimensional generalization of the energy-stress 
tensor and $\Gamma$ is the coupling constant. 
From the theoretical point of view, this yields a theory that 
is covariant under the complete O(4,1) or O(3,2) (depending on which 
signature we choose) as well as under the five dimensional translational 
group. 
The five coordinates are treated equally and the relation to physical 4d 
spacetime has to be found from physical arguments. 

The second way is the induced matter approach. 
In this theory, the 5 dimensional
vacuum equations are splitted in a 4+1 form and interpreted as 4 dimensional 
equations for gravitational fields plus matter fields (see the review article
 \cite{overduin} for instance). Indeed, using 
the metric (7), the equation $G_{AB}=0$  can be written 
\begin{eqnarray}
^{(4)}G_{ik} &=& 
 \frac{\phi_{;i;k}}{\phi}
-\frac{\epsilon}{2\phi^2} \nonumber \\
&& [\ \frac{\dot \phi \dot g_{ik}}{\phi} - \ddot g_{ik}+ g^{lm}\dot g_{im}
\dot g_{lk} - \frac{1}{2} g^{lm}\dot g_{lm}\dot g_{ik} \nonumber \\&& 
 +\frac{1}{4} 
g_{ik}(\dot g^{lm}\dot g_{lm} +(g^{lm}\dot g_{lm})^2)\ ]
\end{eqnarray}

and the additional equation
\beq 
\epsilon\phi \phi^{;i}{;i}= -\frac{1}{4}\dot g^{lm}\dot g_{lm}-\frac{1}{2}
g^{lm}\ddot g_{lm} + \frac{\dot \phi}{2\phi}g^{lm}\dot g_{lm}.
\eeq 
The tensor $ ^{(4)}G_{ik}$ as well as the covariant derivates are formed
with the 4 dimensional part of the metric, $g_{ik}$. The dot means 
derivation with respect to $x^4$. The right hand side of  
(12) is now identified as $\frac{8\pi G}{c^2} T_{ik}$. In this way, 
four dimensional matter arises from five dimensional vacuum. It is claimed 
that these equations recover all the equations of state commonly used in 
astrophysics and in cosmology. Apart from the fact that it would require 
at least some imagination to identify the right hand side of equation (12) 
with the energy-stress tensor of a point-particle for instance, another 
remark is worthlike to be pointed out. $ ^{(4)}G_{ik}$ beeing constructed 
exactly as in general relativity, the 4 dimensional Bianchi identity
$ ^{(4)}G^{ik}_{\ \ ;i}=0$ also holds true. Thus we are left with the 
equation $T^{ik}_{\ \ ;k}=0$ (for this strange $T^{ik}$, the right hand side 
of (12)), which is the equation of motion for the 
(4 dimensional) matter fields. This, however, corresponds, in the 
point-particle case, to the 4 dimensional geodesic equation. 

More generally, $^{(4)}G_{ik}$ being a four dimensional tensor, i.e. 
transforming covariantly under the Poincar\'e group, the same has 
to be true for the right hand side of (12). Thus, we are actually led back 
to a effective four dimensional theory, which is not an extension of 
general relativity, but rather a constraint version, since the 
energy-stress tensor has to be of the particular form (r.h.s. of (12)) 
and in addition, the constraint equation (13) has to be fulfilled. 

Especially, the identification of $x^4$ with $m$ does not seem to 
make much sense in this case. It should however be noticed in favour
of this interpretation that 
if $g_{ik}$ does not depend on $x^4$, the effective 4d  energy-stress
tensor has to be traceless, and thus describes a radiation-like 
equation of state, i.e. massless particles. 

As a conclusion, we remark that in the induced matter approach, we are 
led back to 4 dimensional equations of motion. The physical meaning 
of the five dimensional metric remains unclear and the relation of the 
fifth coordinate to the particle masses is doubtful. 

\subsection{Five dimensional general relativity}

There is still another argument against the induced matter 
approach. The solution (9), which is a vacuum solution of 
the 5d Einstein equations,  suffers from a singularity at the 
spatial origin. This is quite unusual for real vacuum solutions 
(which are wavelike) and should be related to some boundary 
conditions due to the matter distribution. This can be done if we 
include matter fields in the way of equation (11).

For our specific solution (9), 
it is easy to see that $G_{ik} = ^{(4)}\!\!G_{ik}$
and that $g_{ik}$ is just the Schwarzschild solution, which 
satisfies 
\bed ^{(4)}G^{ik} 
= \frac{8\pi G}{c^4}\frac{M_0}{\sqrt{-g^{(4)}}}
\frac{\de x^i}{\de \tau}\frac{\de x^k}{\de t}\ \delta^{(3)}
(\vec x - \vec x(\tau)).
\eed
Thus, in order to write down the equations $G^{AB}=(8\pi \Gamma /c^4) T^{AB}$, 
we have to find a five dimensional tensor $T^{AB}$ whose spacetime components
are of the form 
\bed
T^{ik} = \frac{G}{\Gamma}\ \frac{M_0}{\sqrt{-g}}\frac{\de x^i}{\de \tau}
\frac{\de x^k}{\de t}\ \delta^{(3)}(\vec x-\vec x(\tau)).
\eed

Note that, except from the fact that $g = ^{(4)}\!\!\!\!g$, the right hand 
side of this equation contains the 4 dimensional parameter $\tau$
as well as a 3 dimensional mass density $M_0\ \delta^{(3)}
(\vec x - \vec x(\tau))$. 

If this extension 
can be done in an appropriate way, the equation $T^{AB}_{\ \ ;A}=0$
reduces to the five dimensional geodesic equation (for the particle of 
constant mass $M_0$), i.e. to the 
extremization of the five dimensional line element $\de S$. In this 
case, the five dimensional space has a real physical meaning and 
the deviations from general relativity (due to non-vanishing $\de x^4/\de S$)
can be discussed, even though the identification of $x^4$ and $m$ remains 
still doubtful. 

It is however not straightforward to find the 5 dimensional energy-stress 
tensor that describes a point particle (or some other mass distribution). 
If we require the equations of motion to be 5d geodesics, the action 
has to be of the form 
\bed S_m = \int{q\ g_{AB}u^Au^B}\de S, 
\eed
where $u^A=\frac{\de x^A}{\de S}$ or equivalently, in differential form 
\bed S_m = \int{\mu \ g_{AB}u^A\frac{\de x^B}{\de t} }\  \de^5 x
\eed 
where $q$ is some constant and $\mu = \frac{\de q}{\de V \de x^4}$.
Hence, the Lagrangian density is of the form
\bed \mathcal L_m = 
\frac{\mu}{\sqrt{-g}}\  g_{AB}\frac{\de x^A}{\de S}\frac{\de x^B}{\de t} 
\eed
and the corresponding energy-stress tensor is given by
\bed T^{AB} = \frac{\mu}{\sqrt{-g}}\ \frac{\de x^A}{\de S}
\frac{\de x^B}{\de t}.
\eed
The impact of the introduction of a fifth dimension is best seen in the 
Newtonian limit of the field equations. If we suppose that the 
\textit{matter distribution} $\mu$ is described in the comoving frame, i.e. 
if we set $u^{\alpha}=0,\ \alpha=1,2,3$ and $u^4=0$, and consider 
the first order approximation $g_{00}= 1 + 2 \phi /c^2$, $g_{AB} = \eta_{AB}$ 
else, the equations, up to terms of higher order in $\phi$, reduce to  
\beq 
\Delta \phi \pm \frac{\partial^2 \phi}{\partial (x^4)^2} = 4 \pi \Gamma\  \mu.
\eeq 
The sign on the left hand side again depends on the signature of the 
metric. Note that there are no time derivates, since they are of higher 
order (due to the $c$ that occurs in $x^0 = ct$). Appart from the 
derivates with respect to $x^4$, the main difference from the Newtonian 
field equations lies in the definition of the matter distribution $\mu$. 
As a generalization of the equation $\rho = \de m / \de V$, the 
quantity $\mu$ was defined as a four dimensional density $\mu = \de q / \de V
\de x^4$. This cannot be avoided if we ask for the energy-stress tensor to 
transform covariantly. 

The quantity $q$ thus plays the role of the conserved 
source charge (which in general relativity is $M$) and in the point particle 
case, it is the quantity that characterizes the particle (apart from its 
spin and its electric charge). Since we do not know any other quantities 
that describe a point particle, $q$ should, at least in some \textit{classical}
limit, reduce to the mass. This, in turn, makes it very difficult to interpret
the fifth coordinate as the mass. Indeed, if we try to describe a point 
particle with a definite mass, the density should take the form 
\bed 
\mu = q\ \delta^{(3)}(\vec x - \vec x(\tau))\ \delta(m - m(\tau)),
\eed
where we write $x^4=m$, omitting the  conversion factor for simplicity.
(Note that $\vec x(\tau)$ and $m(\tau)$ are actually constant in the 
frame we have chosen.) But certainly, with this matter distribution, 
the solution will not be 
of the form $\phi = m/r$, because of the singularity not only at 
$r=r(\tau)=0$ but also at $m = m(\tau)=m_0$ in the mass distribution. 
Having in mind however that the Newtonian solution is linear in $m$, 
the  equations (14) take the form
\bed 
\Delta \phi = 4 \pi \Gamma \ \mu,
\eed 
and if we compare this with the Newtonian equation $\Delta \phi = 4\pi G \rho$
we can identify 
\beq \mu = \frac{G}{\Gamma}\ \rho,
\eeq 
or in the particle case
\bed 
\mu = \frac{G}{\Gamma} m_0\ \delta^{(3)}(\vec x -\vec x(\tau)).
\eed
This leads to the correct result, its interpretation is however doubtful. 
If we integrate over 3d space, we find
\bed 
\frac{\de q}{\de m}= \int \mu\ \de^3 x = \frac{G}{\Gamma}\ m_0.
\eed 
The meaning of this equation is that the charge $q$ depends 
\textit{homogenously} 
on the mass coordinate $m$, or in other words, no definite 
mass can be associated 
with the particle. 

We conclude that we cannot interpret $x^4$ as being related to the mass.
In general, we retain the fact that we can find the Newtonian solution 
in the case where $\phi$ does not depend on $x^4$ and if we set $\mu$ as 
in equation (15). This can easily be generalized to the full equations (11)
and the result is again the metric (9), with the difference that we 
can now justify the existence of the singularity at $r=0$. 
In this special case, nothing new 
is gained, except from the additional equation of motion for test 
particles in this metric, $u^4= const$. In particular, this means that 
we have $u_iu^i = const$ as well as $u_Au^A = const$, which could be 
of specific interest in the case of massless particles (light deflection). 
However, for the moment, we don't have neither a theoretical 
interpretation nor an experimental result that allows us to fix the 
additional initial condition $u^4(S = 0)$.

We close this section with some general remarks. One may wonder if the 
conclusions we have got at in both the case of the induced matter approach 
and in the five dimensional theory with additional matter Lagrangian 
are not actually confined to the special solution (9) on which we based 
our discussion. Indeed, even in the spherical symmetric case, there exist 
other solutions (see \cite{gladush} for instance, 
where a generalization of Birkhoff's 
theorem in the context of space-time-mass theory is discussed). The main 
conclusions remain however valid for every solution: In the induced matter 
approach,  the (4d) Bianchi identity leads to 4 dimensional equations of 
motion, and in the theory with additional matter fields, the 3d mass
density $\rho = \de m/\de V$ has to be replaced by a 4d density 
$\mu = \de q/ \de V\de x^4$ (or more generally, a 5 dimensional description
of matter fields is needed). 

The first is of course no problem. It leads 
us back to an effective 4d theory. The physical meaning of the 5d space 
has however to be carefully investigated, especially with respect to 
transformation that mix the spacetime coordinates with the fifth coordinate.
If we take the interpretation from Kaluza-Klein theory to identify the
mixed components with the electromagnetic potential, we have to analyse 
carefully the field equations and the equations of motion in the case 
of a non-compactified fifth dimension. 

As to the second theory, it seems difficult to find a decent five dimensional
description for the matter field. In \textit{classical}, i.e. compactified
Kaluza-Klein theory, the five dimensional quantities can always be 
related to a four dimensional quantity through the integration over 
the compactified dimension. So, for instance, one could refind the mass density
 $\rho$ as $\int{\mu\ \de x^4}$. In our, non-compactified theory, this is 
however not possible, since it leads in general to infinite results 
(this can already be seen from equation (15), whose right hand side 
cannot be integrated in this way). On a more fundamental level, the 
use of equations (11) require a general 5d description of the matter fields,
i.e. a 5d description of particle physics and quantum fields. 

\subsection{Mass as a fifth momentum component}

Consider the special relativistic energy-momentum relation
\beq E^2 -  c^2 \vec p^{\ 2} = m^2 c^4 \eeq
If one intends to introduce variable rest masses through a fifth dimension, 
this equation 
suggests to interpret $m$ not as the fifth coordinate, but ruther as a fifth 
 component of the momentum, i.e. to write 
\beq E^2 - \vec p^2 - m^2 = p_Ap^A = 0,
\eeq 
where for simplicity, we have set $c=1$. Note that the above substitution  
 has been made possible by the fact, that the mass occurs explicitly 
only once in  equation (16)\footnote{Especially, just as in the case of the 
Dirac equation, (16) is not independent of the mass, as are classical
equations
of motion for free particles}. If one repeats the steps that lead from 
(16) to the Dirac equation, the equation (17) yields
\bed i\gamma^A\partial_A\ \psi = 0, \eed
with the corresponding Lagrangian density 
\beq \mathcal L = \bar\psi(i\gamma^A\partial_A)\psi. 
\eeq
In order for the momentum $p_A = i \partial_A$ to satisfy equation (17), 
the matrices $\gamma^A$ have to fulfill the following anticommutation relation:
\bed \{\gamma^A,\ \gamma^B\} = 2\eta^{AB},
\eed
where the signature of the five dimensional Minkowski metric is now clear 
from (17). It is easily seen that the Dirac matrices $\gamma^i$ together 
with the matrix $\gamma^5 = \gamma^0\gamma^1\gamma^2\gamma^3$ will do 
this\footnote{Note that $\gamma^5$ differs by a factor $i$ from the 
usual definition, in order to satisfy $(\gamma^5)^2=-1$. (In the remaining
of this section, the indices $A, B, \dots $ take the values $0,1,2,3,5$.)}. 

Thus, by interpreting $m$ as a fifth momentum component, the four dimensional
Dirac equation is replaced by a five dimensional \textit{massless} Dirac 
equation. In the Lagrangian, the mass term $\bar\psi m \psi$ has been 
replaced by the term $\bar\psi (i\gamma^5\partial_5)\psi$, or in an explicit 
representation (see \cite{kaku} for instance),
\bed m \Rightarrow \left( \begin{array}{cc} -\partial_5 & 0 \\ 0 & \partial_5 
\\ \end{array}
\right). \eed  
 Noether's theorem now yields a conserved current 
\bed J^A = \bar\psi \gamma^A\psi,\ \ \ \partial_AJ^A=0.
\eed 
Integration over 3-dimensional space does, however, not lead to 
a conserved charge. We find instead: 
\bed \frac{\de }{\de t}\ \int{J^0\de^3x} - 
\frac{\de }{\de x^5} \int{J^5\de^3x}=0,
\eed
where we have assumed that $J^0$ does not depend on $x^5$ and vice-versa.  
If we take the usual interpretation for the first term as the time 
derivate of the (unit) charge, the 
second term represents a violation of charge 
conservation.  

As a final remark to the Dirac equation, we note that global U(1) 
gauge invariance leads in the same manner as in the 4 dimensional case 
to a 5 dimensional gauge theory through the extension\footnote{To obtain 
a locally gauge invariant theory, $e$
  has  to be a constant. It can thus not be identified with $\int\de^3xJ^0$.} 
$\partial_A \Rightarrow \partial_A + ieA_A$. We will try to identify $A_5$ 
with the gravitational 
potential.   

We know that in the 4 dimensional case, to the Lagrangian (18) corresponds 
a classical Lagrangian $L= u_iu^i$. In the 5 dimensional case, this leads
to a problem, because we cannot simply divide equation (16) by $m^2$
without breaking the covariance of the theory. (This is the reason why we 
began this section with the Dirac equation and not with the classical theory.)
To solve this problem, we have to reintroduce a constant mass $m_0$ which 
we interpret as the restmass far from gravitational fields and we write
\bed m = m_0 u^5. \eed
Thus, $u^5$ can be interpreted as the variation of the \textit{unit} mass. 
Further, we define $u^i = p^i/m_0$. We can now write down the following 
Lagrangian for free particle motion:
\beq
L = \eta_{AB} u^Au^B = u_iu^i - (u^5)^2.
\eeq
Note that due to (17), this Lagrangian is actually zero (in analogy 
to (18), a 5d Dirac equation for \textit{massless} particles). This Lagrangian
now satisfies the properties we have requested at the end of section II.B, 
namely that in the absence of the gravitational field, the mass has to 
be found to be constant. Indeed, the solution of the equations of motion 
are just $u^A = const$, and $u^5$ is set to one because of $m=m_0u^5$. 

Now, we proceed introducing a gravitational potential. If we suppose that 
the fifth component of the gauge potential $A_A$ is responsible for 
the gravitational interaction, we can write 
\beq 
L = \frac{1}{2} u_Au^A -  \phi u^5,
\eeq 
where $A_5 \sim \phi$ (the coupling constants have been absorbed into $\phi$). 
Since $L$ is of the form $u_Au^A/2 - A_Au^A$, in analogy to the EM case, 
$u^Au_A$ is still constant (and hence equal to zero). 
If we suppose that $\phi$ is a function of $r$ only, we have, 
in addition,  
the following constants of motion (in spherical coordinates $r,\psi,\theta$):
\bedd l &=& r^2 \dot \psi \\ 
      d &=& \dot t \\
      \epsilon &=& u^5 + \phi,
\eedd   
where we have used the Euler-Lagrange equation of the coordinate $\theta$ 
to set $\theta = \pi/2$ (planar motion). The dot denotes differentiation with
respect to the (5 dimensional) curve parameter. With these constants, the 
relation $u_Au^A =0$ can be written in the form
\beq \dot r^2 = \de^2 - \epsilon^2 + 2 \epsilon \phi - \frac{l^2}{r^2} 
- \phi^2. 
\eeq 
In a parameter independent form, introducing $\rho = 1/r$, this can 
be written as
\beq (\frac{\de \rho}{\de \psi})^2  = \frac{d^2-\epsilon^2}{l^2}+
\frac{2\phi\epsilon}{l^2}-\frac{1}{r^2}- \frac{\phi^2}{l^2}.
\eeq
It is tempting to set $\epsilon = 1$, because in the limit $r \to \infty$, 
$u^5$ should tend to 1. This, however, cannot be done, because we 
consider planetary motion, i.e. confined motion, and the limit $r \to \infty$ 
has no physical  meaning in this case. (In analogy, the constant $a=\dot t(1-
2R/r)$ that arises in general relativity (Schwarzschild metric) cannot 
be set to one. This would lead only to geodesics with 
zero (newtonian) energy.) On the other 
hand, our constant $d = \dot t$ can of course be set to one, it is just 
a rescalation of the curve parameter. 

Neglecting the higher order term $-\phi^2$, the above equation can be 
identified with the Newtonian equation of motion if we set $\phi = - \phi_N$, 
with $\phi_N = -GM/r$ the Newtonian potential. To compare the higher 
order corrections, we remind that the spherically symmetric post-newtonian 
extension of general relativity, to an order that describes all the
observational  effects, leads to the equation
\beq (\frac{\de \rho}{\de \psi})^2  = \frac{a^2-1}{l^2}-
\frac{(4a^2-2)\phi_N}{l^2}-\frac{1}{r^2}+ \frac{6a^2\phi_N^2}{l^2}.
\eeq 
The main difference lies in the sign of the correction term $\sim \phi_N$, 
which is opposite to the one in (22). It is possible to \textit{force}
equation (22) into the form (23) if we extend our potential $\phi$ 
as $\phi = \alpha + \beta \phi_N + \gamma \phi_N^2 + \dots $. 
The Newtonian limit
leads, as we have seen, to $\beta = -1$. The positive factor infront of 
the $\phi_N^2$ term  requires that $\gamma > 0$. This, however, means 
that $\phi$, and hence also $u^5$, which is related to $\phi$ 
through $\epsilon = u^5 + \phi$, will annulate at some values $r$ and 
can be represented (at least up to the considered order) with periodic 
functions ($u^5 \sim \sin(a\phi_N) \sim \sin(b/r)$\dots) which does not 
fit the interpretation of $u^5$ as the variation of the unit mass.     
Note that if the last sign in (22) were opposite, we could bring the 
equation into the form (23) with a relation of the form $\phi -\epsilon 
\sim u^5 \sim \exp(-\phi_N)$ which 
would be similar to the mass definition in section (II.B), i.e. the
interpretation of the mass as the opposite of the gravitational energy.
This would however require a different signature of the 5d metric. 

The conclusion of these considerations is that the fifth component of 
the potential $A_A$ cannot be, in this way, interpreted as gravitational
potential. The question remains what it represents instead. This can 
only be answered after we have propperly defined a conserved charge (if this 
is possible at all). The problem is actually the same as the one discussed
in section (IV.E). It is not possible, in a straightforward way, to describe
matter using four dimensional \textit{charges} in the context of a 
five dimensional theory. 

On the other hand, even in the induced matter approach, we were dealing
with a Lagrangian of the form $L = u_Au^A$, and to this, some five 
dimensional Dirac equation should correspond, with or without  
mass term. So, even if this approach did not lead to fundamental problems
on the level of general relativity, it may do so on the quantum theory 
level. 

Actually, the considerations of the last section are very close to the 
induced matter approach, since we have interpreted the matter term 
$\bar\psi m \psi$ as fifth component of the Dirac equation, and achieved 
a five dimensional \textit{massless} equation. Note, however, that 
Noether's theorem applied on the Lagrangian (18) yields the following
energy-stress tensor
\bed T^{AB} = i \bar\psi \gamma^A \partial ^B \psi, \eed
which is traceless (because of the Dirac equation), but not zero.

To the second approach of 
section (IV.E), where five dimensional matter is explicitly introduced, 
would correspond
a Dirac Lagrangian of the form 
\bed \mathcal L = \bar\psi(i\gamma^A\partial_A-m)\psi, 
\eed 
 where $m$ is now a constant and the meaning of the fifth component 
remains open. It yields the same energy-stress tensor (which is 
not traceless in this case, because of the mass term in the Dirac 
equation).  

\section{Conclusions}

The considerations of the last section, but also those of the Newtonian 
case, lead us to conclude that we cannot use the variation of the 
mass (or more generally, an additional dimension) to describe the 
gravitational interaction, i.e. to replace curved spacetime by a higher 
dimensional flat space. So, as we have pointed out earlier, the 
variation of the mass has to be considered as an additional degree of 
freedom. Further, we came to the conclusion, that in the absence of 
gravitational fields, the masses are found to be constant. In other 
words, the additional degree of freedom is directly related to the 
gravitational fields. These properties are all included in a five 
dimensional general relativity theory (see metric (7)), if we 
interpret the mass as 
the fifth momentum component. The classical limit is found in the case 
were $g_{ik}$ in (7) is independent of $x^4$ (we can then remove the 
field $\phi$ through the transformation $\de x^4 \rightarrow \phi \de x^4$).

The unsolved problem is the question of the matter description. In the 
induced matter approach, the fact that $g_{ik}$ does not depend on $x^4$ 
means that we are dealing with 4d vacuum solutions (see (12)), which 
is unsatisfactory for the description of the field of a planet or of a 
point particle. More realistic descriptions should thus include an $x^4$ 
dependency or a five dimensional energy-stress tensor, which leads however
to other difficulties. 

A general problem of five dimensional theories is the fact, that  
the conseved current density does not lead to a conserved charge 
(or mass). A relation of the form $\partial_A J^A = 0$ leads 
(if $\partial_0 J^5= \partial_5 J^0 =0$)
to 
\bed \frac{\de }{\de t}\ \int{J^0\de^3x} - 
\frac{\de }{\de x^5} \int{J^5\de^3x}=0.
\eed
In order to get a charge that is conserved throughout time, we have 
to integrate not over 3 dimensional space, but over a four dimensional 
space, i.e. we have 
\bed \frac{\de }{\de t}\int{J^0 \de^3x \de x^5} = 0. \eed    
This does, however, not correspond to the notion of the total charge 
contained in three dimensional space. Especially, such a relation makes
it rather difficult to suppose a direct relation between $m$ and $x^5$ 
(or $p^5$). 

These problems suggest that before we introduce an additional dimension 
into a gravitational theory,  we have to establish a well defined 
five dimensional elementary particle description.   

Finally, we remark that all these problems are absent in scalar-tensor
theories of the kind of Brans-Dicke theory. This is due to the fact that 
the only field that couples to matter in this approach is the metric, 
whereas the scalar field only contributes to the energy density. Thus, 
the scalar field reacts on matter only indirectly through its contribution
to spacetime curvature, i.e. its influence on the metric. Hence, 
a consistent Dirac 
equation in the framework of this theory just consists in the usual 
generally covariant form of this equation (with the help of the 
spin-connection). This observation is an additional hint to the fact 
that we should regard, in this theory, the coupling constant as  
variable, not 
the rest mass.

\end{document}